# Waveform measurement of charge- and spin-density wave packets in a Tomonaga–Luttinger liquid

M. Hashisaka[1], N. Hiyama[1], T. Akiho[2], K. Muraki[2], and T. Fujisawa[1]

[1] Department of Physics, Tokyo Institute of Technology, 2-12-1-H81 Ookayama, Meguro, Tokyo 152-8551, Japan,
[2] NTT Basic Research Laboratories, NTT Corporation, 3-1 Morinosato-Wakamiya, Atsugi, Kanagawa 243-0198, Japan.

In contrast to a free-electron system, a Tomonaga–Luttinger (TL) liquid in a one-dimensional (1D) electron system hosts charge and spin excitations as independent entities[1-4]. When an electron wave packet is injected into a TL liquid, it transforms into wave packets carrying either charge or spin that propagate at different group velocities and move away from each other. This process, known as spin-charge separation, is the hallmark of TL physics. While the existence of these TL eigenmodes has been identified in momentum- or frequency-resolved measurements[5-13], their waveforms, which are a direct manifestation of 1D electron dynamics, have been long awaited to be measured. In this study, we present a time-domain measurement for the spin-charge-separation process in an asymmetric chiral TL liquid comprising quantum Hall (QH) edge channels[12-15]. We measure the waveforms of both charge and spin excitations by combining a spin filter with a time-resolved charge detector. Spatial separation of charge and spin-wave packets over a distance exceeding 200 μm was confirmed. In addition, we found that the 1D electron dynamics can be controlled by tuning the electric environment. These experimental results provide fundamental information about non-equilibrium phenomena in actual 1D electron systems.

Non-equilibrium dynamics in 1D electron systems are described in terms of the excitation of TL eigenmodes[5-17]. For example, TL eigenmodes are responsible for the rich variety of non-equilibrium phenomena in QH edge channels, including heat transport[18-21], decoherence in interferometers[22,23], interference of high-frequency excitations[12] and noise generation[13,14,24]. Here, we demonstrate a time-domain waveform measurement for the eigenmodes in co-propagating edge channels of the QH state at filling factor $\nu$ = 2, a prototypical system for the study of spin-charge separation[12-14]. The measured waveforms provide essential information about the quantitative aspects of non-equilibrium behaviours in QH edge channels.

A schematic of the measurement is shown in Fig. 1. We study the electron dynamics in the co-propagating spin-up and -down channels, labelled 'target channels' in the figure. Inter-channel interaction yields two eigenmodes, i.e. charge and spin modes, that comprise charge excitations in both channels. The charge mode has the same charge polarity in the two channels, while the spin mode has the opposite polarity. Our aim is to obtain time-domain waveforms of these eigenmodes' wave packets. This is possible by decomposing the excitations into independent spin-up and -down charges in each channel and measuring the charges using a time-resolved charge detector[25]. First, an initial charge-mode wave packet [(i) in Fig. 1] is excited and fed to the spin filter SF$_{in}$ placed at the entrance of the target channels. A spin-up (Fig. 1a) or -down (Fig. 1b) charge packet (ii) is injected through SF$_{in}$ into the channels by filtering out the other spin component. The spin-polarized charge packet splits into charge- and spin-mode packets (iii and iv), which are spatially separated after propagation. Each separated packet is decomposed by the second spin filter SF$_{det}$ placed in front of the time-resolved charge detector. Only the spin-up charge components are measured by the detector to construct the objective waveforms of the TL eigenmodes.

The 1D electron dynamics in this system are formulated by the wave equation for the spin-up and -down density excitations

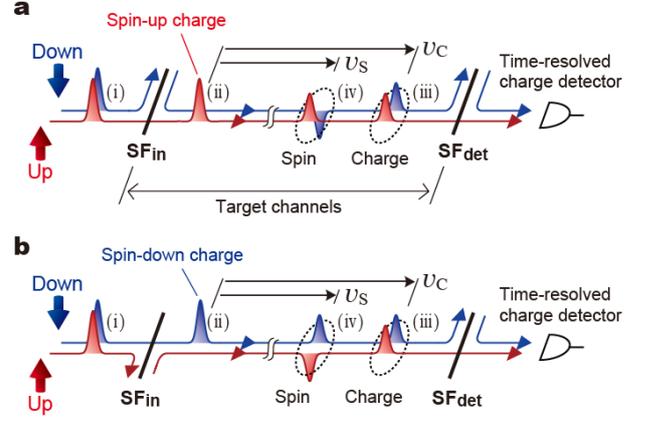

**Figure 1 Principle of the experiment. a** Spin-up charge injection into the target channels through SF$_{in}$. The spin-polarized charge packet splits into the charge- and spin-mode wave packets with different group velocities $v_C$ and $v_S$, respectively. The spin-up charge components of these packets are extracted through SF$_{det}$ and are detected by the time-resolved charge detector. A pair of same-charge-polarity peaks should be detected. **b** Spin-down charge injection. Peak and dip structure of opposite charge polarity should be detected in this case.

$\rho_{\uparrow,\downarrow}(x, t)$[15,26,27]

$$\frac{\partial}{\partial t}\begin{pmatrix}\rho_\uparrow\\\rho_\downarrow\end{pmatrix}=-\mathbf{U}\frac{\partial}{\partial x}\begin{pmatrix}\rho_\uparrow\\\rho_\downarrow\end{pmatrix}=-\begin{pmatrix}v_\uparrow & U_X\\U_X & v_\downarrow\end{pmatrix}\frac{\partial}{\partial x}\begin{pmatrix}\rho_\uparrow\\\rho_\downarrow\end{pmatrix},$$

where $U_X$ is the inter-channel interaction and $v_\uparrow$ ($v_\downarrow$) is the group velocity renormalized by the intra-channel interaction in the spin-up (down) channel. While conventional TL theory assumes $v_\uparrow = v_\downarrow$, we consider a more general case wherein they can differ. By diagonalising the matrix $\mathbf{U}$, we obtain transport eigenmodes: the charge mode $\boldsymbol{\rho}_C = (\rho_{C\uparrow}, \rho_{C\downarrow}) = (\cos\theta, \sin\theta)$ with velocity $v_C$ and the spin mode $\boldsymbol{\rho}_S = (\rho_{S\uparrow}, \rho_{S\downarrow}) = (\sin\theta, -\cos\theta)$ with velocity $v_S$ ($\leq v_C$)[12-14]. The mixing angle $\theta$ ($0 \leq \theta \leq \pi/2$) represents the asymmetry between the two channels. Pure spin-charge separation ($\theta = \pi/4$) occurs only for the symmetric case with $v_\uparrow = v_\downarrow$.

In the present experiment, a spin-polarized packet ($d(t)$, 0) or (0, $d(t)$) is induced at $x = 0$ to excite both $\boldsymbol{\rho}_C$ and $\boldsymbol{\rho}_S$ wave packets. Here, $d(t)$ is the real-time charge distribution function centred at $t = 0$. The time evolution of the system is described as a superposition of the two modes

$$\boldsymbol{\rho}(x,t) = \cos\theta\,\boldsymbol{\rho}_C d(t - x/v_C) + \sin\theta\,\boldsymbol{\rho}_S d(t - x/v_S).$$

We measure the spin-up component $\rho_\uparrow(L, t)$ using the detector placed at $x = L$. In an ideal TL liquid, $\boldsymbol{\rho}_C$ and $\boldsymbol{\rho}_S$ propagate without attenuation; thus, $v_C$, $v_S$ and $\theta$ (and thus $v_\uparrow$, $v_\downarrow$ and $U_X$) are directly extracted from the obtained waveforms. In an actual 1D electron system, attenuation of these eigenmodes results in distortion of the waveforms from which deviation from the ideal TL behaviour can be evaluated.

Figure 2a shows the experimental setup and optical micrograph of the device. The measurement was performed at

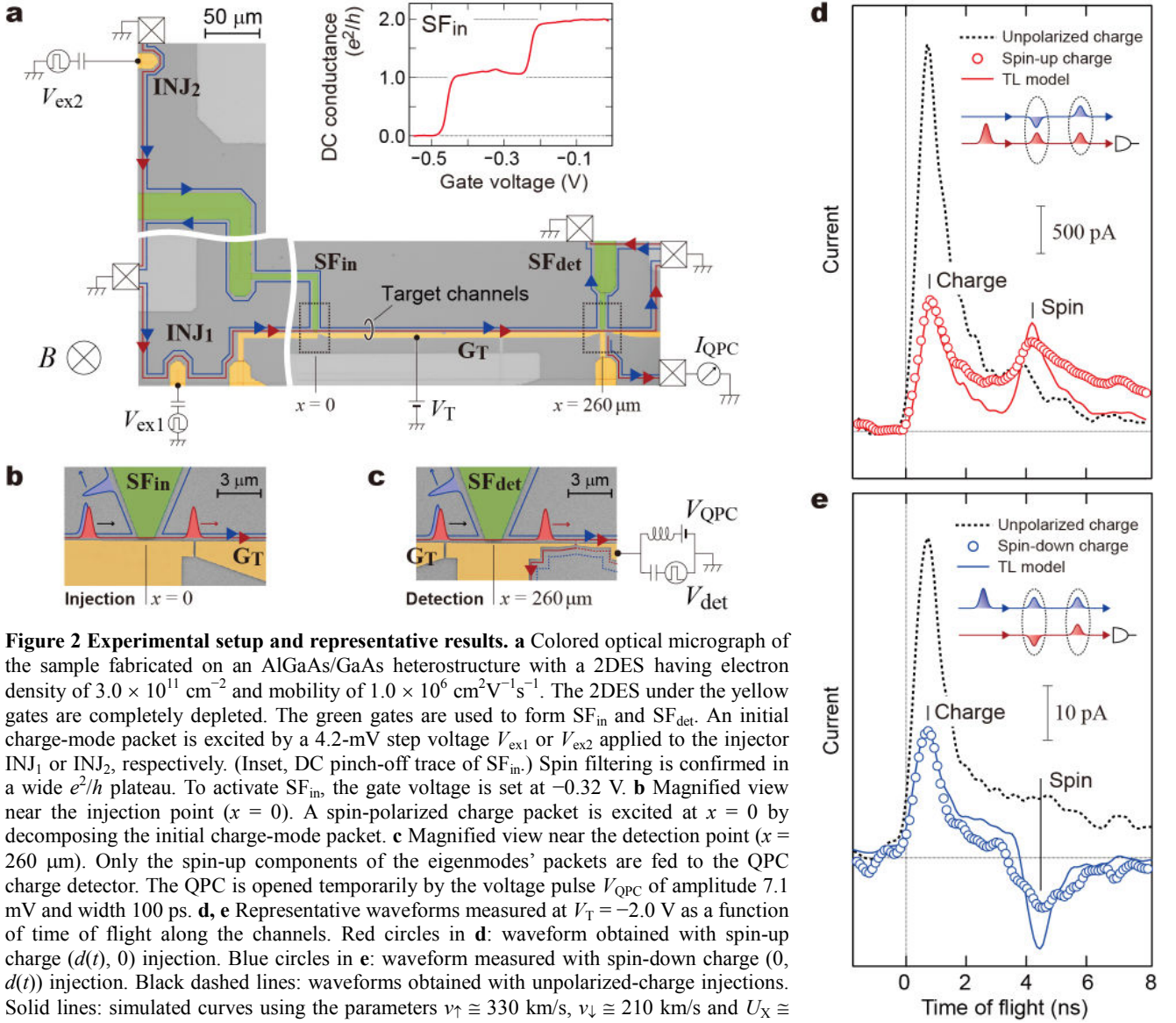

**Figure 2 Experimental setup and representative results. a** Colored optical micrograph of the sample fabricated on an AlGaAs/GaAs heterostructure with a 2DES having electron density of $3.0 \times 10^{11}$ cm$^{-2}$ and mobility of $1.0 \times 10^{6}$ cm$^2$V$^{-1}$s$^{-1}$. The 2DES under the yellow gates are completely depleted. The green gates are used to form SF$_{in}$ and SF$_{det}$. An initial charge-mode packet is excited by a 4.2-mV step voltage $V_{ex1}$ or $V_{ex2}$ applied to the injector INJ$_1$ or INJ$_2$, respectively. (Inset, DC pinch-off trace of SF$_{in}$.) Spin filtering is confirmed in a wide $e^2/h$ plateau. To activate SF$_{in}$, the gate voltage is set at −0.32 V. **b** Magnified view near the injection point ($x = 0$). A spin-polarized charge packet is excited at $x = 0$ by decomposing the initial charge-mode packet. **c** Magnified view near the detection point ($x = 260$ μm). Only the spin-up components of the eigenmodes' packets are fed to the QPC charge detector. The QPC is opened temporarily by the voltage pulse $V_{QPC}$ of amplitude 7.1 mV and width 100 ps. **d, e** Representative waveforms measured at $V_T = -2.0$ V as a function of time of flight along the channels. Red circles in **d**: waveform obtained with spin-up charge ($d(t)$, 0) injection. Blue circles in **e**: waveform measured with spin-down charge (0, $d(t)$) injection. Black dashed lines: waveforms obtained with unpolarized-charge injections. Solid lines: simulated curves using the parameters $v_\uparrow \cong 330$ km/s, $v_\downarrow \cong 210$ km/s and $U_X \cong 200$ km/s. The data in Figs. **d** and **e** are measured with different cooldowns.

320 mK in a magnetic field $B = 6.5$ T ($\nu = 2$) applied perpendicular to the 2D electron system (2DES). Spin-up and -down edge channels (indicated by red and blue lines, respectively) are formed along the energized surface gates (green and yellow patterns). We focus on the 260-μm target channels along gate G$_T$, which are formed by applying gate voltage $V_T$ to completely deplete 2DES under the gate.

An initial charge-mode packet is excited at the injector INJ$_1$ or INJ$_2$ by applying step voltage $V_{ex1}$ or $V_{ex2}$, respectively[17]. The spin filters SF$_{in}$ and SF$_{det}$ comprise local QH regions at filling factor $\nu = 1$ formed under the top gates where only the spin-up channels travel through the gated regions (inset in Fig. 2a). When the initial packet is fed from INJ$_1$ (INJ$_2$), a spin-up (-down) charge packet is injected into the target channels through SF$_{in}$ (Fig. 2b). The spin-polarized charge packet splits into the charge- and spin-mode packets, which should be measured at different timings by the spin- and time-resolved charge detector comprising SF$_{det}$ and the quantum point contact (QPC) (Fig. 2c)[17,25].

The open circles in Figs. 2d and 2e represent the central results of this paper, which are the waveforms obtained when a spin-up and -down charge packet is injected, respectively. We observe the first peaks at the time of flight (approximately 0.5 ns) as well as an additional peak in Fig. 2d and a dip in Fig. 2e at approximately 4 ns. These features correspond to those expected for spin-charge separation, as shown in the inset. The distinct spin dependence of the current polarity allows us to unambiguously identify the first and second features as the charge and spin modes, respectively. This is reconfirmed by comparing the data with the results of the control experiments, where $SF_{in}$ is deactivated so that both spin components pass through it. As the injected spin-unpolarised charge packet is an eigenmode in this case, a single peak is observed, indicating no spin-charge separation (black dashed lines in Figs. 2d and 2e). The peak positions coincide with those of the first charge-mode peaks observed in the spin-polarized charge injection case.

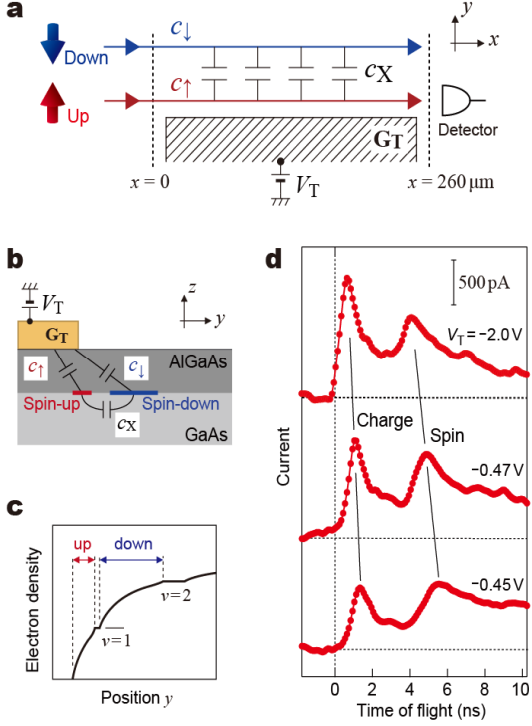
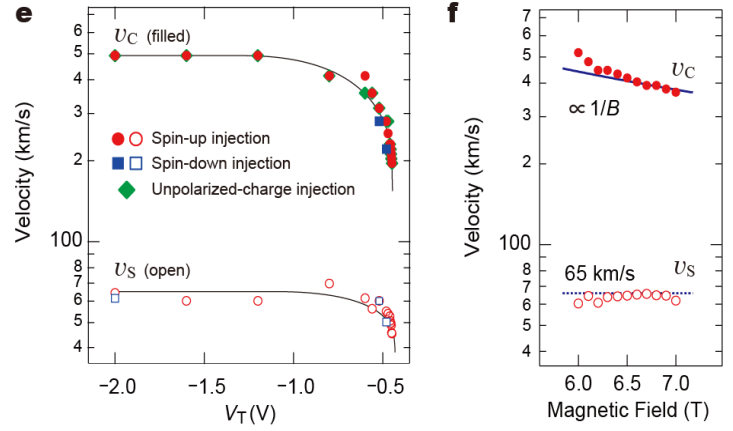

**Figure 3 Tuning the transport properties of TL eigenmodes.**
**a** Distributed-element circuit model of target channels. **b** Schematic cross section of the target channels. **c** Schematic of the electron-density profile at the edge of the 2DES. Compressible strips form the spin-up and -down edge channels. **d** Representative waveforms with spin-up charge injection measured at $B = 6.5$ T and at $V_T = -2.0$, $-0.47$, and $-0.45$ V. **e** $V_T$ dependences of $v_C$ (filled) and $v_S$ (open symbols). Solid black lines are visual guides. **f** Magnetic field $B$ dependences of $v_C$ and $v_S$ measured at $V_T = -2.0$ V. The velocity $v_C$ varies with $B$ following the $1/B$ curve (solid line), while $v_S$ remains nearly constant at ~65 km/s (dotted line).

The clearly resolved peak–peak (peak–dip) structure is a manifestation of the long-range ballistic transport of the TL eigenmodes. This implies that the co-propagating channels are well described by the chiral TL-liquid theory. From the arrival times of the charge and spin modes, we obtain $v_C \cong 480$ km/s and $v_S \cong 65$ km/s. The ratio of the charge- and spin-mode peak heights $I_C$ and $I_S$, respectively, measured in case of the spin-up charge injection (Fig. 2d) provides the mixing angle $\theta \cong 0.2\pi$. From these parameters, we obtain the TL parameters as $v_\uparrow \cong 330$ km/s, $v_\downarrow \cong 210$ km/s and $U_X \cong 200$ km/s.

The ballistic transport properties cause large spatial separation of charge and spin excitations. The distance $D$ between the excitations is estimated from $v_C$ and $v_S$. For example, $D \cong 225$ μm at the charge-mode detection timing (0.54 ns). Thus, we successfully demonstrate a single spin-charge-separation process.

The solid lines in Figs. 2d and 2e represent the waveforms simulated using the obtained $v_C$, $v_S$ and $\theta$ values without considering attenuation. In these calculations, $d(t)$ is determined assuming $d(t - L/v_C) \propto \rho_{0,\uparrow}(L, t)$, where $\rho_{0,\uparrow}(L, t)$ is the spin-up component of $\boldsymbol{\rho}(x,t)$ measured with the unpolarised-charge injections (dashed lines). The main features of the experiment and simulations agree quite well. A small deviation in the shape of the spin-mode peaks (approximately 20% difference in the peak heights) suggests the presence of attenuation caused by the inter-channel spin-flip tunnelling (Supplementary information).

In the rest of this paper, we discuss tuning of the 1D electron dynamics. The transport properties of the TL eigenmodes depend on the surrounding environment, which screens the interactions in the channels. The screening strengths are represented by distributed electrochemical capacitances $c_\uparrow$, $c_\downarrow$ and $c_X$ (Fig. 3a)[21,28]. Figures 3b and 3c show a schematic cross section of the channels and the electron-density profile at a filling factor slightly greater than $\nu = 2$, respectively. Compressible strips separated by the incompressible strip form the co-propagating channels[29]. The widths and positions of these strips vary with experimental parameters, resulting in the tuning of $c_\uparrow$, $c_\downarrow$ and $c_X$, and thus the TL parameters.

Figure 3d shows the waveforms for spin-up charge injection measured at several $V_T$ values below the depletion voltage $V_{T,dep} = -0.44$ V. While the two-peak structures are observed in all traces, the shape and positions of the peaks vary with $V_T$. The group velocities $v_C$ and $v_S$ extracted from the peak positions are summarised in Fig. 3e. The velocities increase monotonically with decreasing $V_T$, showing a steep enhancement at $-0.55$ V $< V_T < -0.45$ V. This reflects the fact that the edge channels move far apart from the gate metal with decreasing $V_T$, thereby reducing $c_\uparrow$ and $c_\downarrow$. The steep enhancement near $V_{T,dep}$ reflects the edge configurations that rapidly change in this $V_T$ range.

The TL parameters can also be tuned by the magnetic field $B$. Figure 3f shows the $B$ dependences of $v_C$ and $v_S$ near $v = 2$. While $v_C$ varies following the curve $v \propto 1/B$ (solid line) corresponding to the theory of edge magnetoplasmon transport[30], $v_S$ remains nearly constant at 65 km/s. This behaviour is also explained by considering the $B$ dependences of the channel configurations (Supplementary information). In this manner, the TL behaviours can be modified by the environmental parameters. The variety of TL eigenmodes' properties results in a variety of non-equilibrium phenomena in QH systems.

In summary, we presented a simple and effective experimental scheme to obtain the waveforms of charge- and spin-mode wave packets in a chiral TL liquid. The dynamics of TL eigenmodes, including excitation, propagation and attenuation, are directly evaluated from the measured waveforms. While the measurement was performed on the QH edge channels in this study, the concept of combining spin filters with a time-resolved charge-detection technique can be widely applied to various 1D electron systems.

**Acknowledgements**
The authors thank H. Kamata, N. Kumada and Y. Tokura for their beneficial discussions. This work was supported by Grants-in-Aid for Scientific Research (26103508, 15H05854, 26247051, 16H06009).

**Author contributions** M.H. and T.F. designed and supervised this study. N.H. and M.H. performed the experiment and analysed the data. T.A. and K.M. grew the wafer. M.H. wrote the manuscript with help from T.F. and K.M. All authors discussed the results and commented on the manuscript.

**Author information**
Reprints and permissions information is available at www.nature.com/reprints. The authors declare no competing financial interests. Readers are welcome to comment on the paper. Correspondence and requests should be addressed to M.H. (hashisaka@phys.titech.ac.jp) and T.F. (fujisawa@phys.titech.ac.jp).

**Methods**
**Sample fabrication.** We fabricated the sample in a standard AlGaAs/GaAs heterostructure with a 2DES located 95 nm below the surface. The sample was patterned using photolithography for chemical etching and coarse metalized structures, as well as e-beam lithography for fine gate structures. The ohmic contacts shown by the white squares with cross marks in Fig. 2a were formed by alloying Au-Ge-Ni onto the surface.
**Time-resolved charge detection.** This technique was performed by applying two RF voltages, i.e. step voltage to excite an initial charge packet and square pulse voltage for temporary opening of the QPC. These rf voltages generate a charge flow through the QPC when the detection pulse synchronises to the arrival of a charge excitation at the QPC. We measured the current $I_{QPC}$ transmitted through the QPC as a function of the time delay $\tau$ of the detection pulse from the excitation voltage under successive applications of these voltages at a repetition frequency of 25 MHz.

In this experiment, the time origin of the waveform measurement is defined as the moment of the spin-polarized charge excitation at $x = 0$. We estimated the moment in the $I_{QPC}$ trace by the following procedure. First, we defined a straight channel from $INJ_1$ to the detector by applying −2.0 V to the gates (total length of the channels is $L_{total} = 560$ μm). Second, an $I_{QPC}$ trace was measured with the charge-mode packet injection at $INJ_1$ without activating $SF_{in}$. The arrival time $\tau_C$ of the packet at the detector was evaluated from the peak position. Then, the excitation timing $\tau_0$ with charge injection at $INJ_1$ was calculated as $\tau_0 = \tau_C \times (L_{total} - L)/L_{total}$, where $L = 260$ μm is the length of the target channels. With charge injection at $INJ_2$, we performed similar measurement to estimate arrival time $\tau_C'$ and calculated the excitation timing as $\tau_0' = \tau_C' - L/v_C$. In this way, we determined the time origins of the waveform measurements as $\tau_0$ and $\tau_0'$.